\RequirePackage{etex} 
\documentclass[runningheads,twocolumn,a4paper,10pt]{llncs}
\usepackage[maxnames=6,firstinits=true,doi=false,url=true,isbn=false]{biblatex}
\usepackage{etex}
\reserveinserts{28}
\pdfoutput=1
\usepackage[a4paper, total={7in, 10in}]{geometry}
\usepackage[linesnumbered,lined,boxed,commentsnumbered,ruled]{algorithm2e}

\SetAlFnt{\small}
\SetAlCapFnt{\small}
\SetAlCapNameFnt{\small}
\SetAlCapHSkip{0pt}
\IncMargin{-\parindent}
\usepackage{ textcomp }
\usepackage{ comment }
\usepackage[color=yellow!60,textsize=footnotesize,obeyDraft,draft]{todonotes}
\usepackage{colortbl}
\usepackage{booktabs}
\usepackage{capt-of}
\usepackage{makecell}
\usepackage{multirow}
\usepackage{listings}
\usepackage{graphicx}
\usepackage{amsmath}
\usepackage[font=footnotesize,labelfont=bf]{subcaption}
\usepackage[hidelinks]{hyperref}
\usepackage[capitalize]{cleveref}
\usepackage{tikz}
\usepackage{pgfplots}
\usepackage[para,online,flushleft]{threeparttable}
\usepackage{wasysym}
\usepackage{amssymb}
\usetikzlibrary{arrows}
\usepackage{breakcites}
\usepackage{array}
\usepackage{color}
\usepackage{balance} 
\usepackage{lastpage}
\usepackage{dblfloatfix}
\usepackage{authblk}
\usepackage{siunitx}
\usepackage{xurl} 
\usepackage[shortlabels]{enumitem}

\usepackage{float}
\floatstyle{plaintop}
\restylefloat{table}
\newcommand{\pathToOtherFiles}{other}

\begin{document} 

\title{Safety Validation of Autonomous Vehicles using Assertion Checking}
\titlerunning{Safety Validation of Autonomous Vehicles... }
\authorrunning{Harper, Chance, Ghobrial et al.}

\author{Christopher Harper\inst{1}, Greg Chance\inst{2}, Abanoub Ghobrial\inst{2}, Saquib Alam\inst{1}, Tony Pipe\inst{1}, Kerstin Eder\inst{2}}

\institute{University of the West of England, Bristol, UK \and University of Bristol, Bristol, UK}
\tocauthor{Authors' Instructions}

\maketitle
\let\thefootnote\relax\footnotetext{
\textit{Statements about authorship contribution.}
Christopher Harper (e-mail: chris.harper@brl.ac.uk),
Saquib Alam (e-mail:saquib764@gmail.com), 
and
Tony Pipe (e-mail: tony.pipe@brl.ac.uk), 
are with the University of the West of England, Frenchay, Coldharbour Ln, Bristol, BS34 8QZ, United Kingdom.
Greg Chance (e-mail: greg.chance@bristol.ac.uk), 
Abanoub Ghobrial (e-mail: abanoub.ghobrial@bristol.ac.uk), 
and 
Kerstin Eder (e-mail: kerstin.eder@bristol.ac.uk) 
are with the Trustworthy Systems Lab, Department of Computer Science, University of Bristol, Merchant Ventures Building, Woodland Road,  Bristol, BS8 1UQ, United Kingdom. }

\makeatletter
\renewcommand\subsubsection{\@startsection{subsubsection}{3}{\z@}%
                       {-18\p@ \@plus -4\p@ \@minus -4\p@}%
                       {4\p@ \@plus 2\p@ \@minus 2\p@}%
                       {\normalfont\normalsize\bfseries\boldmath
                        \rightskip=\z@ \@plus 8em\pretolerance=10000 }}
\makeatother

\textbf{\textit{Abstract}--Safety and mission performance validation of autonomous vehicles (AVs) is a major challenge. In this paper we describe a methodology for constructing and applying assertion checks to validate the behaviour of an AV operating either in simulation or in the real world. We have identified a taxonomy of assertion types and the general format of their specification, and we have developed procedures for translating driving codes of practice to yield formal logical expressions that can be monitored automatically by computer, either by direct translation or by physical modelling. 
We have developed examples of assertions derived from the UK Highway Code (UKHC), as an example of a code of practice. 
%
%
We illustrate the approach with an example of assertion checking for vehicle overtaking, using a geospatial information system in an SQL database for validation and performance assessment.
We present initial simulation and runtime monitoring 
experiments that apply assertions relevant in this overtaking scenario together with an analysis of the safety and mission performance characteristics measured.
%
}


\section{Introduction}

Verification and validation are important to earning trust and gaining confidence in the safety of autonomous systems (AS) such as autonomous vehicles (AVs). Safety validation of any system can (and should) be performed by a variety of measures, including analysis and inspection of designs and their implementation. In this paper we focus on validation by system testing, and in particular on using assertions for a vehicle overtaking scenario  both in simulation and in runtime monitoring. 
%

We are investigating the use of assertions to validate autonomous vehicle behaviour against driving codes of conduct ("rules of the road"). 
This is challenging not only because driving rules are often vague and may be conflicting, but also because of the contextual and common sense knowledge required to interpret these rules into appropriate driving actions. 
Field testing, often the traditional approach to validating systems in their target environments, is likely to be prohibitively expensive for autonomous vehicles
, if a high degree of coverage or statistical confidence is required. This has led to much interest and research into the use of driving simulators\footnote{Examples include \url{https://www.rfpro.com} and \url{http://carla.org}.} for validation. 

We focus specifically on methodologies that incorporate the classification, measurement and evaluation of a system's situated behaviour. One such approach is based on the specification of situated behavioural properties as \emph{assertions}, logical expressions that can be applied 
to monitor the behaviour of the system under test and to report any property violations. 
These properties relate to the observable driving behaviour of road vehicles. They characterise what any vehicle operator, autonomous or human, 
%
%
may be expected or legally required to achieve on public roads. 
%
%
In many countries, these requirements are usually defined in legally sanctioned codes of practice for human drivers.
%
We have used the UK Highway Code~\cite{highwayCode} (UKHC) as our working example, 
to investigate the how to develop validation assertions for properties such as safety and mission performance. 


%
We demonstrate how to systematically transform these rules into assertions. To achieve this, the driving rules must be translated from human-readable format to machine-readable expressions that can be automatically monitored. 
Although we select the UKHC as 
our reference 
for pragmatic reasons, it could equally be replaced with other metrics pertaining to functionality or even social convention. We review some related work in Section~\ref{Related_work}, in which other potential sources for 
sets of assertions 
have been identified.
%
%

%
We illustrate the approach
with two examples of 
checking lane changing behaviour
during an overtaking scenario. We adopt the terms scene, scenario and situation from~\cite{Ulbrich2015} throughout this paper. 
%

Evaluation of AV behaviour against codes of driving conduct is emerging as an essential aspect AV 
certification, as set out in the recent UK Law Commission report~\cite{law_commission_UK}. 
%
Assertion 
checking may be used as evidence to show functional safety compliance against national regulations or codes of practice, the general safety argument being that if an AV satisfies all the rules that may be legally expected of a human driver, then its behaviour is comparable to that of human drivers and therefore should be acceptable, or at least legal. It should be noted that acceptance may therefore be somewhat context-specific, against the code of practice for each country, and what passes under one legal code may not be acceptable in another. 
We anticipate that there may need to be extensive harmonization of national driving codes to avoid having to re-certify vehicles for each new national market and improve the cost-effectiveness of certification processes, but this is outside of the paper's scope.

The need to measure the compliance of an AV's behaviour with legal codes of practice may go beyond the initial design validation or certification stage as performed in simulation and (to some extent) road tests. It may become necessary to measure such compliance during vehicle operation, for example to establish that the vehicle's behaviour is correct (or otherwise) in the event of an incident that requires subsequent investigation. If an AV is involved in an incident, but it can be shown by assertion-based assessment during operation that it was adhering to all relevant legal standards of driving behaviour, then this may have an impact on any liability or insurance-based compensation that may be due as a consequence of the incident. A detailed discussion of the rationale and requirements for explainability~\cite{rosenfeld2019explainability} of AV decision-making in post-incident investigation is beyond the scope of this paper.


This paper makes the following contributions. 
We develop and present a taxonomy of assertions that covers the externally observable behaviours of vehicles (although in principle this is applicable to measurements of any system). 
We present a novel technique for deriving, systematically formalising and encoding assertions from the UKHC and applying them in two modes: (i) within a 
vehicle simulator at the vehicle design validation stage, and (ii) within a runtime monitoring system during vehicle operation.
In the first mode of use, we encode the UKHC assertions as SQL queries and run them within a PostgreSQL database with a PostGIS extension.\footnote{Refer to \url{https://www.postgresql.org/} and \url{http://postgis.net}.} In this mode, the assertions can make use of any data stream available, passed into the database records.
In the second mode of use, the principal requirement is that data about the AV, other road users and the adjacent environment with respect to the test vehicle is sufficient to evaluate the assertions during runtime. The same PostgreSQL-based database as used in the simulator can therefore be installed as embedded software, but its data stream will be sourced from real vehicle sensors instead of a simulator. This data stream need not be of a high fidelity, but rather should have ``just enough'' detail~\cite{Koopman2018} to allow the assertions to be evaluated with good confidence.

In the following, related work is reviewed in Section~\ref{Related_work}. Section~\ref{generic_architecture} explains the structure of the PostgreSQL database and how it integrates with the underlying simulator or runtime monitoring environment. Section~\ref{Use_of_assertions} discusses the  methodology of deriving assertions from the UKHC, and the principles of using 
assertion checking
both in simulations and for runtime monitoring. Section~\ref{Experimental_scenario} introduces the overtaking scenario we have used to demonstrate the application of assertion 
checking
and describes two case studies to illustrate the two options, simulation-based testing and runtime monitoring. The results from the case studies are presented and analysed. In Section~\ref{discussion} we discuss some important aspects of assertion-based monitoring, which have emerged from the experimental work, in particular the use of assertions for performance monitoring as well as for safety, and the observation that in many assertions the intention of ego-vehicle agents may also need to be monitored to demonstrate compliance with certain driving rules. Finally, in Section~\ref{conclusion} we draw conclusions and suggest directions for future work.

\section{Related Work}
\label{Related_work}

Challenges exist to formalise human-defined traffic rules that can then be used as the ground truth 
basis
to verify the correctness of a vehicle's behaviour. 
Prakken et al.~\cite{lawabidingstudy} discusses some of these issues and argue in favour of the need for formalisation of traffic rules. Rizaldi et al.~\cite{acountability} use the formalisation as a means to resolve accountability issues in an event of an accident. To the best of our knowledge, most work~\cite{acountability, esterle, rizaldi, alves} advocates the use of Higher-Order Logic (HOL) to codify a wide range of traffic rules into a formal language. The authors of~\cite{acountability, esterle, rizaldi, alves} use Linear Temporal Logics (LTL) to construct a set of equations for various manoeuvres to ensure road safety. To construct these equations, a formal model of each lane and lane marking is required. The authors of~\cite{rizaldi} used \textit{lanelets}~\cite{lanelets2014} to build these lane models. While many works use the German or Vienna convention~\cite{vienna} on road traffic, in~\cite{alves} a road junction rule from the UKHC is used to demonstrate their methodology. Some researchers~\cite{sqlhuang, sqlgueffaz} have used temporal logic queries to formulate the HOL, using a variety of query languages like SQL and RQL. 

DeCastro et al.~\cite{decastro2018counterexample} guide an agent to learn probabilistic models, via a learning-based algorithm, constrained by safety contracts which are in essence traffic rules. However, the major challenges for using learning algorithms are completeness and verification of the solutions. Such systems are essentially a black box, and it is extremely difficult to validate their correctness. Another major problem is what reward or cost function should be used to train the model. One approach can be to have an 
assertion-checking system
providing rewards for correct behaviour, 
leading
back to developing traditional assertion-based rules using the relevant highway code.

The Responsibility Sensitive Safety (RSS) framework developed by MobilEye~\cite{RSS_Shalev_Shwartz2017, RSS2_Koopman2019} proposes mathematical models for mimicking the human subjective decision making as an effort to provide safe driving behaviour by AVs. On the other hand, legal codes such as the UKHC often define the interactions required with particular features of their respective road environments (for example, junctions, roundabouts, or crossings). Our work is intended to address this latter domain, although a complete database will need to incorporate sets of assertions of both categories. 

All the works discussed earlier are limited to detecting the occurrence of some undesirable events while driving. However, such events are not equal in severity~\cite{sinha}. Myers et al.~\cite{myers} propose a framework to access autonomous driving systems using two types of scoring rules: prescriptive and risk-based. To the best of our understanding, the prescriptive scoring mechanism captures the capability of the AV to follow traffic rules while the risk-based scoring system assesses the behaviour of the AV in rare events where undesirable outcomes are unavoidable, and it is expected of the AV to exhibit behaviour that would minimise the risk. Developing a generic system to score risks in different scenarios is challenging and requires a separate study of its own. Therefore, risk-based assessment is not part of this paper.

We use SQL to build the assertions, which is highly expressive~\cite{sqllibkin} and can be integrated with powerful geospatial object handling technology (PostGIS). This provides a natural environment to write logical checks that can be performed on static and dynamic objects in a geographical context. 
%

\section{System Architecture for Assertion-based Monitoring} \label{generic_architecture}

The evaluation of assertions requires an environment to capture or generate vehicle data, and apply the assertions to interpret the driving situation. This section presents the generic components and software architecture we have deployed to support assertion-based monitoring either in simulation or during runtime.

\subsection{Assertion Monitoring Architecture} \label{generic_sim_system}

\begin{figure}
    \centering
    \includegraphics[width=0.98\linewidth]{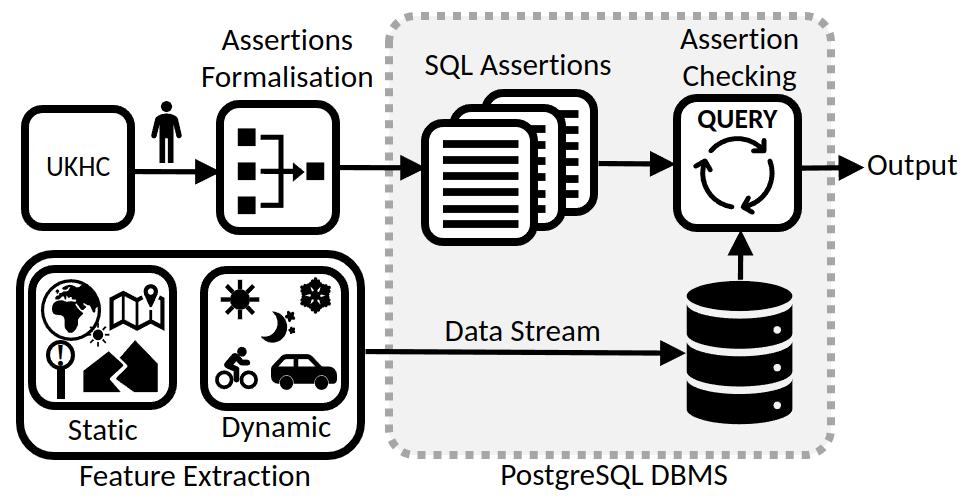}
    \caption{Process for interpreting and monitoring assertions.}
    \label{fig:SimulatorArchitecture}
\end{figure}

Figure~\ref{fig:SimulatorArchitecture} shows the process developed to formalise and monitor assertions and interpret the driving situation. The process starts with the reference used to assess the driving situation, in this case the UKHC. Following this, there is a human-interpreted stage (human icon) where human-readable rules are converted into formal assertions (\emph{Assertion Formalisation}) that are machine-readable. This forms one of two inputs to the 
PostgreSQL database 
, which performs
the automated assertion checking. The second input to the 
assertion checking system
is the \emph{Data Stream} of extracted features that is observing the driving scenario, which may be from simulation or from live sensor data. The \emph{Feature Extraction} process will capture relevant information from the driving scenario, including \emph{Static} elements such as the road network and \emph{Dynamic} elements such as vehicle position and orientation. Records are captured from the \emph{Data Stream} and stored on the PostgreSQL DBMS server at each available time step.
%
Automated assertion checks are executed within the DBMS, where \emph{SQL Assertions} are stored which express the formalisations of the UKHC in SQL. At each time step the \emph{Assertion Checking} process invokes a query engine that evaluates the relevant assertions against the stored records and generates a pass or a fail verdict as an \emph{Output}.

\subsection{Geospatial Database for Assertion Checking} \label{geospatial_database}
To develop assertions derived from UKHC rules, we need a formal language into which they can be translated for automatic checking. Two aspects are important to the choice. First, as we discuss later, the rules and guidelines in the UKHC are written for human readers not automatic systems. They consist of sets of discrete rules applicable to specific aspects of road user behaviour, which must be taken collectively, and not necessarily always in a particular order, to form the legally required profile of behaviour. This suggests that a declarative programming language, such as Prolog or SQL, may offer the required expressive power for encoding rules.
Second, to process the data produced during driving, we seek languages that handle large data sets naturally, where the assertions constitute formal operations on datasets corresponding to the logical properties we extract from the UKHC rules. 
Taking these two requirements into consideration, SQL emerges as the most suitable programming language. SQL has been demonstrated formally~\cite{sqllibkin} to be sufficiently expressive to be able to encode related problems such as state or graph reachability as well as all standard problems in relational algebra, albeit with the use of some of its later extensions such as recursive operators. Safety properties can be defined in terms of reachability~\cite{guiochet2015, masson2019} or (in-)stability~\cite{harper2005, xue2020} of desirable (goal) and undesirable (hazardous) situated states. For these reasons, we have selected SQL-based database technology as the framework for our assertion 
checking system.

The technical challenge then becomes how to compare the data captured against the assertions. We argue that a geospatial relational database is well suited for this purpose. The vehicle state can be easily stored and accessed. Additional insights, such as dynamic properties, can also be derived from this base information. 
%
Thus, we have implemented Assertion Checking as a PostgreSQL database with a PostGIS geospatial information system extension, which provides a comprehensive library of SQL-native spatial measurement functions. We chose these particular software products because they are open-source, and PostgreSQL implements powerful, high-performance table searching algorithms.
Real applications will involve the use of large data sets, so the performance of the underlying database engine is an essential factor in practice. 

\begin{figure}
    \centering
    \includegraphics[width=8.5cm]{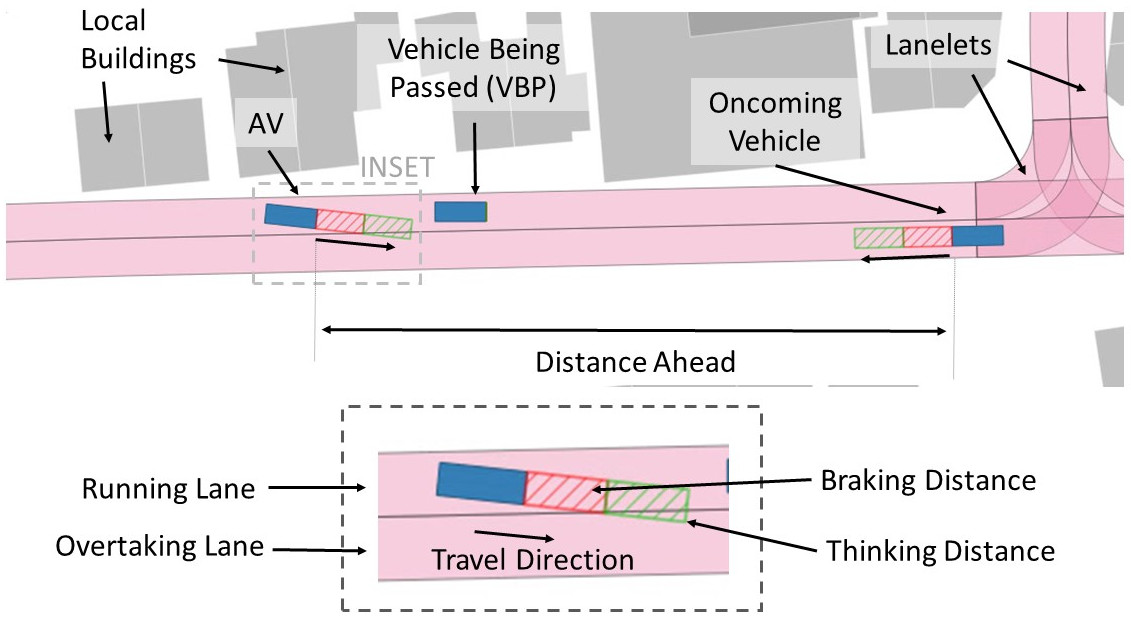}
    \caption{Database view of the overtaking scenario.}
    \label{fig:assertion_database_annotated}
\end{figure}

The use of PostGIS introduces an extensive library of data types and method functions suited to measurement of basic physical (geometric) relationships. For the overtaking scenario in this paper, functions measuring distance between geometric objects (the bounding-box shapes of vehicles on the road), and partial or total containment of the vehicle within road lane geometries, are of particular use to assess vehicle positions during the overtaking manoeuvre. 
PostGIS functionality can also be used to generate more complex geometric dynamic data derived from the basic state of the vehicle or other agents. An example of this is shown in Figure~\ref{fig:assertion_database_annotated}, in which two polygons for \textit{thinking distance} (inset, green hashed area) and \textit{braking distance} (red hashed area), which are parameters defined in the UKHC (Rule 126), can be generated as a function of the basic forward velocity data of the vehicle. Figure~\ref{fig:assertion_database_annotated} shows a projection of these derived geometries onto the associated street map. 
Thus, new abstractions can be developed by deriving new data sets from the original (captured) data. These can be manipulated effectively as a new variable (SQL attribute) appended to the original data set, and referenced by queries in the same way.
%


This is an essential principle of our methodology. Procedures and queries can be developed to generate abstractions directly related to the concepts expressed in the rules of the UKHC. 
Where such concepts are used in multiple rules, the corresponding abstractions form a library of common queries or operations, which can be incorporated into the DBMS. 
%
%
%
%

To illustrate the functionality of the PostgreSQL DBMS server, Figure~\ref{fig:assertion_database_annotated} shows the operating situation associated with the overtaking scenario described in Section~\ref{Experimental_scenario}. 
In this example the AV is on the left of the figure overtaking a parked vehicle requiring a lane change. The distance to the oncoming vehicle is termed the \emph{Distance Ahead} (DA). 
The figure shows an urban scene depicting a section of road network divided into \emph{lanelets} denoting legal division between driving lanes and any buildings in the area. 
Following one of the clauses in Rule 162 in UKHC we develop an assertion to determine whether the distance ahead is sufficient for a safe overtake manoeuvre (this assertion is called \emph{Safe Distance Ahead} (SDA)). 
%

The assertion monitor performs a check (as defined in UKHC Rule 162) that a safe distance ahead exists between the AV and the nearest vehicle in an oncoming lane, labelled the `Oncoming Vehicle' (OV) as the AV starts to overtake the `Vehicle Being Passed' (VBP). The start time of the overtake is defined as the time that the AV first begins to enter the adjacent lane (Overtaking Lane) of the road, at which point it must commit to the manoeuvre or else abort and return to driving in the initial lane (the Running Lane).

In the case of assertions related to overtaking, it is necessary to determine whether the overtaking vehicle wholly or partly occupies a given road lane, since overtaking manoeuvres require it to move at least partially into an adjacent lane and back. Therefore, it is necessary for the database to include lane geometries (polygons), as shown in Figure~\ref{fig:assertion_database_annotated}, in order to allow checking of vehicle occupancy of lanes by testing for the overlap between vehicle and lane objects. 
The overtake manoeuvre ends when the vehicle is entirely within the running lane once more after having passed the VBP. Lane geometry objects (shape or bounding box polygons) are therefore an essential element of the assertions related to overtaking.

\section{Assertion Development Methodology}\label{Use_of_assertions}
In this section we first identify the types of assertions that may exist. We then discuss how to develop assertions from the natural language statements of the UKHC. We also present how to use assertions, both in simulation and during runtime.

\begin{figure}[!b]
    \centering
    \includegraphics[width=0.98\linewidth]{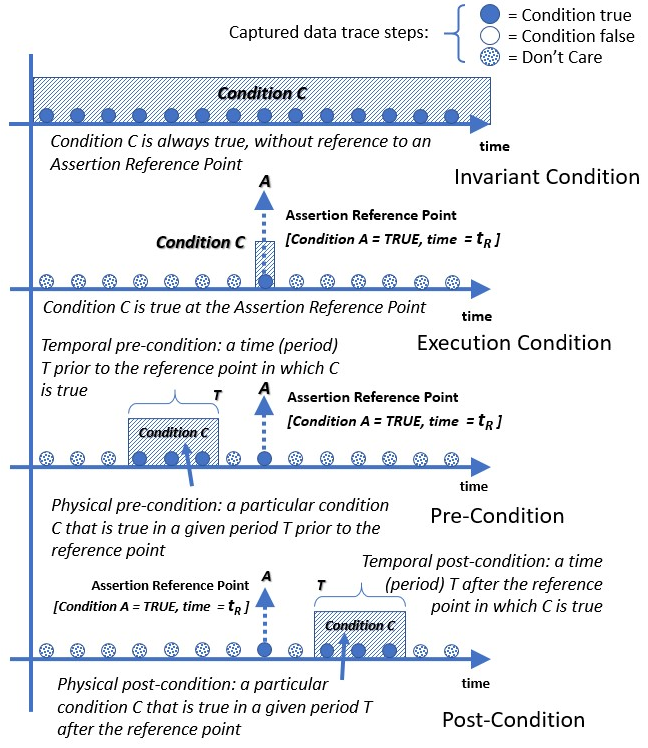}
    \caption{Types of assertions.}
    \label{fig:categories_of_assertions}
\end{figure}

\subsection{Types of Assertions} \label{assertion_categories}
Where UKHC rules define objectively measurable driving behaviour, the assertions to encode them tend to fall into one of four principal types. Except for invariants, assertions are defined with respect to some \emph{assertion reference point} - a particular step of a captured data trace, defined by some reference condition which holds at a specified time step. The assertion is then defined as some spatio-temporal condition relative to the reference point, as shown in Figure~\ref{fig:categories_of_assertions}.

The four types of assertions are:
\begin{itemize}
    \item Invariant Condition: A condition that must be satisfied at all time steps within the captured data trace.
    \item Execution Condition: A condition that must be satisfied at the assertion reference point.
    \item Pre-condition: A spatio-temporal phenomenon that must exist in the steps preceding the assertion reference point. 
    \item Post-condition: A spatio-temporal phenomenon that must exist in the steps following the assertion reference point.
\end{itemize}

Pre- and post-conditions can be of one of two sub-types:
\begin{itemize}
	\item \textit{Temporal}: A time period preceding or following the assertion reference point, within which a specified condition holds. This is the temporal component of the overall spatio-temporal condition, e.g.:
	\begin{itemize}
		\item \textit{Temporal pre-condition}: as a vehicle crosses a highway exit lane boundary (reference point) checking that the vehicle's indicators have been set continuously in a preceding period (e.g.\ 5s) (i.e.\ checking to detect unduly late or intermittent indications).
		\item\textit{Temporal post-condition}: a highway exit assertion may check that if a vehicle starts indicating to leave at the exit lane, then it does enter the exit lane within a given time period (i.e.\ checking to detect unnecessarily early indications).
	\end{itemize}
	
	\item \textit{Physical}: A physical (e.g.\ spatial or state) condition that must be satisfied at a specified time(s) before or after the assertion reference point. This is the physical component of the overall spatio-temporal condition, e.g.:
	\begin{itemize}
		\item \textit{Physical pre-condition}: a check that, as a vehicle passes the exit lane boundary, its indicators were set at the last highway exit sign before the exit (i.e.\ checking to detect omitted indications).
		\item \textit{Physical post-condition}: a check that if a vehicle is indicating to leave in the approach to an exit lane, then it does in fact leave the lane (i.e.\ checking to detect false or spurious indications).
	\end{itemize} 
\end{itemize}

\subsection{Assertion Applicability and Triggering} \label{assertion_triggers}
Identifying which assertions are relevant in a given situation is important to manage the performance of assertion checking, which is  
particularly critical for runtime monitoring, and to ensure the output is meaningful. For example, assertions specifically for highway driving are not applicable in urban or rural road environments. 
Similarly, different assertions checking for mutually exclusive conditions will be guaranteed to generate errors unless the inappropriate assertions are filtered out. For example, checking that an AV is braking smoothly will inherently fail in a situation where it is required to accelerate. If assertions of this kind are not selected correctly, they will generate nuisance failure results.


Discrimination 
of assertions can be achieved by specifying `environment conditions' based on the type of \textit{Operating Design Domain (ODD)} in which the assertion is relevant, e.g.\ highway versus urban or rural environments.\footnote{This is similar to typical practice in computer hardware design verification~\cite{tao2009}, where an assertion may take the form ``under $<$environment conditions$>$ if $<$trigger$>$ then $<$expectation$>$", where \textit{$<$environment conditions$>$} may be ``we are not in RESET mode" and \textit{$<$trigger$>$} may be ``we receive a request" and \textit{$<$expectation$>$}  may be ``an acknowledge signal is driven high in the next X clock cycles".} 
In addition, encoding the assertion type will enable the assertion checking process (see Figure~\ref{fig:SimulatorArchitecture}) to check only assertions of a given type.
%


\subsection{Deriving Machine-Oriented Assertions from Human-Oriented Rules}
\label{interp_nat_lang_UKHC_rules}

The UKHC comprises several hundred rules and guidelines, applicable to different categories of road user 
(e.g.\ pedestrians, cyclists, cars, trucks and buses) 
and covering a number of different aspects of road use. 
The UKHC rules are written entirely for human readers, and seek to offer guidance in common sense terms that any reader should understand clearly. For example, two rules of central interest to this paper relate to overtaking manoeuvres:

\vspace{2mm}
\noindent\underline{Rule 162}    
    \textbf{Before overtaking} you should make sure
    \begin{itemize}
        \item the road is sufficiently clear ahead
        \item road users are not beginning to overtake you
        \item there is a suitable gap in front of the road user you plan to overtake.
    \end{itemize}
	
    
\noindent\underline{Rule 163}
    \textbf{Overtake only} when it is safe and legal to do so. You should:
    \begin{itemize}
        \item not get too close to the vehicle you intend to overtake
        \item use your mirrors, signal when it is safe to do so, take a quick sideways glance if necessary into the blind spot area and then start to move out
        \item not assume that you can simply follow a vehicle ahead which is overtaking; there may only be enough room for one vehicle
        \item \emph{move quickly past the vehicle} you are overtaking, once you have started to overtake. \emph{Allow plenty of room. Move back to the left as soon as you can but do not cut in}
        \item take extra care at night and in poor visibility when it is harder to judge speed and distance
        \item give way to oncoming vehicles before passing parked vehicles or other obstructions on your side of the road
        \item only overtake on the left if the vehicle in front is signalling to turn right and there is room to do so
        \item stay in your lane if traffic is moving slowly in queues. If the queue on your right is moving more slowly than you are, you may pass on the left
        \item give motorcyclists, cyclists and horse riders at least as much room as you would when overtaking a car (see Rules 211 to 215).
    \end{itemize}

This example illustrates several characteristics found frequently in UKHC rules: %
\begin{enumerate}[i, labelindent=0pt]
\item Rules are often expressed in a second-person tense, offering advice directly to the reader. Converting them into assertions that are in effect performed (measured) from an external third-person perspective requires changes that may transform the logic of the rule to some extent. 
\item The logical sense of the safety property expression (i.e. whether the rule expresses a pass or a fail condition) varies depending on the concept being expressed. Since the general guidance for direct conversion of rules is to keep as close as possible to the top level natural language, the logical sense of the assertions will follow suit. This contrasts with typical practice for their use in conventional computer programming, where the standard practice is to encode an assertion such that a pass condition permits a program to continue running normally, and a fail causes an exception to be triggered.
\item Being written at a natural human-readable level of abstraction, the rules and guidelines rely greatly on the reader having sufficient background knowledge and capability to resolve their generalized constraints to the point where actions can be selected that satisfy them. This abstraction also interferes with the ability to validate the assertions by any means other than manual design review; formalization of the assertions may require all the hidden complexity to be reintroduced. 
\item Many clauses of UKHC rules offer advice to drivers about their internal decision making, often in the form of constraints. The third bullet-item of Rule 163 is a typical example. These clauses cannot be measured externally without some form of communication by the driving agent (for example, an AV) of the results of its decision-making processes. To date we have not attempted to establish assertions of this kind, as there are no standards or conventions by which an AV might explain and communicate its actions. However, we note that this is an interesting direction of future work (see Section \ref{discussion}). 
\end{enumerate}

\subsection{Methods for Deriving Assertions}\label{deriving_assertions}
Deriving assertions from the human-readable UKHC 
requires that it be interpreted in a consistent manner. 
We present two approaches 
for developing assertions from the UKHC: direct translation into logic, and the use of modelling.

\vspace{2mm}
\subsubsection{\textbf{Method 1}: Direct translation into logic}
\label{direct_translation}

This approach involves converting the natural language text into a logical predicate. We developed a procedure containing the following steps: 

\noindent $i$) \emph{Identify and extract a given UKHC rule clause.} 
    \begin{itemize}
    	\item A separate assertion will be required for every distinct subject-verb-object clause in the text; a given UKHC rule could easily require in the order of 5-10 separate assertions to be fully covered;
    	\item the clause must define a \emph{testable} requirement - some UKHC rules are (at least at present) not testable. [We discuss this issue further in Section \ref{discussion}.]
    \end{itemize} 
    \noindent $ii$) \emph{Write a natural language hypothesis that captures the safety property of the rule.} 
    \begin{itemize}
    	\item A useful way to do this is to phrase the natural language hypothesis interrogatively, i.e. as a question.
    	\item Many rules are written imperatively, as orders or advice for road users to follow. The hypothesis is straightforward: were the conditions of the instruction(s) satisfied? 
    \end{itemize}
      
    \noindent $iii$)  \emph{Write a logical statement that reflects accurately the assertion hypothesis}
    \begin{itemize}
    	\item Note that the perspective of the hypothesis may need to be transformed to a `third-person' (external) viewpoint.
    	\item Write the assertion logic as closely as possible to the language of the assertion hypothesis, keeping the same high-level terms.
    \end{itemize}
   
    \noindent $iv$) \emph{Verify assertion logic by manual design review.} 
    \begin{itemize}
    	\item As previously mentioned, pragmatically this is the only method available. But if the assertion logic is directly comparable to the  natural language of the assertion hypothesis, this step should be reasonably self-evident.
    \end{itemize}
    

As an example of the direct translation method, consider the first clause from UKHC Rule 163, mentioned previously. The procedure could be applied as follows:

\noindent $i$) \emph{UKHC rule clause:}
	    \begin{quote}
			\textit{You should not get too close to the vehicle you intend to overtake.}
		\end{quote}
\noindent $ii$) \emph{Assertion Hypothesis:}
        \begin{quote}
        	The hypothesis question is: \textit{As the AV begins to cross the road centre line (start of the overtake manoeuvre), is the separation distance between AV and VBP is less than [a pre-determined distance]?} 
        	
        	The specific separation distance value will have to be specified by some external requirement, as the rule is not specific as to what is acceptable or safe in this context. For the purposes of this example, we declare that the acceptable distance is equal to the sum of the 'thinking distance' and 'braking distance' (as shown in \ref{fig:assertion_database_annotated}) of the AV, as this would allow the AV to decelerate without collision risk if an emergency occurred as it began its overtake manoeuvre.
        \end{quote}
\noindent $iii$) \emph{Assertion Statement (Predicate logic):}

The assertion concept, of testing whether the separation between AV and VBP is always at least equal to the sum of the AV's thinking distance and braking distance at the moment it crosses the road centre line, is an Execution Condition whose elements are defined as follows:
	\begin{itemize}
		\item \emph{Assertion Reference Point(s)}: all time steps where the AV bounding-box geometry overlaps with the road centre line;
		\item \emph{Assertion Condition}: distance between AV and VBP is greater than the sum of the thinking and braking distances (this will be referred to as the `danger space' of the AV).
	\end{itemize}
 
%
This can be converted into an SQL query of a form similar to the fragment shown below:
 \begin{lstlisting}[
            language=SQL,
            showspaces=false,
            basicstyle=\ttfamily,
            numbers=none,
            numberstyle=\tiny,
            commentstyle=\color{gray}]
 WITH danger_space  
     AS /* sum of thinking
               & braking distances */        
 SELECT (time) ASC LIMIT 1
 FROM simulation_data AS sd
 WHERE
  ST_Overlap(AV.geom::geography, 
             /*road centreline*/) AND
  ST_Distance(AV.geom::geography, 
                  VBP.geom::geography)
                      <= danger_space;
\end{lstlisting}

Although the above example is simple enough not to require their use, key concepts of the assertion (for example road centre lines) can be implemented as procedural functions stored in the PostgreSQL database.

We anticipate that the more complex assertion types, such as spatio-temporal Pre- or post-conditions, will require stored procedural functions whereas invariant and execution conditions will probably be simple enough to be handled by straightforward queries.  
Typically, assertion queries will make use of PostGIS relational operators (such as ST\_Distance or ST\_Overlap) and functions for object geometries, whereas the library functions will typically use PostGIS geometry constructors to set up the geometric objects for the assertions as variables derived from the captured simulation data. 
A hierarchical organization of queries and procedural functions (from Assertions to CAV DBMS library functions to PostGIS functions to captured simulation data) emerges from this approach.

\vspace{2mm}
\subsubsection{\textbf{Method 2}: Model-based Analysis} 
\label{model_based_analysis}

While many UKHC rules can be translated directly into logical conditions or constraints expressible as simple Boolean predicates, some describe more complex situations that may require more explicit modelling to identify the assertion hypothesis and logical expression. We require no particular constraints on the modelling methodology used, except that it needs to produce a well formed logical expression deriving the essential parameter of the assertion hypothesis from the data stream.

As a worked example, we present an assertion we developed from Rule 162, which we applied experimentally. 
The first two steps of this method are the same as for Method 1: identify the UKHC clause and the associated assertion hypothesis:

\noindent $i$) \emph{UKHC rule clause:}
		\begin{quote}
			\textit{Before overtaking you should make sure the road is sufficiently clear ahead.}
		\end{quote}
\noindent $ii$) \emph{Assertion Hypothesis:}
	\begin{quote}
		The hypothesis question is:
		
		 \textit{Is the distance between the AV and OV sufficiently large to complete the overtake manoeuvre, at the moment the AV starts it (i.e.\ as it crosses the road centre line into the overtaking lane)?}
	\end{quote}

While the UKHC rule clause seems straightforward, the natural language actually encapsulates a complex judgement that a driver must make at the moment they begin an overtake manoeuvre. The driver must judge whether the distance between their vehicle and any oncoming vehicle in the adjacent road lane is sufficient to allow the manoeuvre to be completed safely. This is contingent on the (relative) speeds of the driver's vehicle (the AV), VBP and OV, line-of-sight distances to limits of view or obscuring obstacles such as bends in the road, blind summits, or road-side buildings, as well as the style of manoeuvre that the driver intends to perform - an `aggressive' overtake manoeuvre can probably be completed in less distance than a `relaxed' one~\cite{decastro2018counterexample, tkachenko2018line}.\footnote{The internal plans and intentions of an AV may not be available to the external validation system, so the definition of the assertions may need to make default assumptions about its intentions, or make a `worst case' test that checks against the most conservative case that can be assumed. For example, using the case of overtaking, the assertion could assume that the AV was making a `relaxed' overtake manoeuvre, in which case an assertion might register a safety violation for anything less than large separation distances between the AV and oncoming vehicle at the start of the manoeuvre, even though the manoeuvre could probably be completed safely if a more urgent profile were performed.} There is a complex relationship between all the elements of this assertion that is quantitative in nature and hence an algebraic formula must be developed that expresses the assertion hypothesis. Such expressions must be developed by modelling the situation.


In the case of this Rule 162 example, it was necessary to develop a complete model of the overtaking manoeuvre in order to produce an expression for the \emph{Safe Distance Ahead} (SDA), which calculates how far away the Oncoming Vehicle needs to be from the AV for an overtaking manoeuvre to be conducted without risk of collision. The analysis decomposed the overtaking manoeuvre into several stages: Pulling Out (into the oncoming lane), Passing the VBP, and Cutting In (to the original running lane). A formula was developed for the overall distance required to complete the manoeuvre without violating safety constraints, such as impingement upon vehicle thinking or braking distance (see Figure~\ref{fig:assertion_database_annotated}) or clearance distances around the VBP. A diagram of the full analysis model, and the derivation of the formula for Safe Distance Ahead, is provided in Appendix A. 




The analysis identifies several key parameters that characterise the overtaking manoeuvre. These include:
\begin{itemize}
    \item \emph{Stopping distance} (SD) is defined as the sum of the `thinking distance' and `braking distance' of the vehicle (assuming it has a human driver). 
    A table of the typical stopping distances for travelling speeds ($v$) between 20~mph - 70~mph is provided with Rule 126 in the UKHC.
    Equation~\ref{DS_equation_rule126} is a regression formula derived from this table to estimate the stopping distance (in metres) 
     given the vehicle speed (in mph).
    The coefficients are: $a=0.300$, $b=0.058$, $c=-0.011$ and $d=0.015$.
    
\begin{equation} 
\label{DS_equation_rule126}
\text{SD} = \underbrace{av }_{\substack{\text{Thinking} \\ \text{Distance}}} + \underbrace{b + cv + dv^2}_{\substack{\text{Braking} \\ \text{Distance}}}
\end{equation}

    \item \emph{Danger Space} (DS) is an area projected forward in the direction of travel with a length equal to the stopping distance and a width equal to that of the vehicle. The overall formula for SDA (given in Appendix A) incorporates the DS of the OV.
    %
    SD varies with vehicle speed, making the DS a function of speed, therefore intuitively as the speed increases the DS ahead of the vehicle will project further forward.
    
    \item \emph{Pull-out Clearance} distance is the minimum safe separation that must be maintained as the AV pulls out into the oncoming lane (and carries its own constraints (and assertions) as defined in the first bullet-point of Rule 163.
    
    \item \emph{Pull-out Angle} ($\beta$) is the steering angle taken by the AV as it moves into the oncoming lane.
    
    \item The \emph{Cut-in Clearance} distance is the minimum required separation distance between AV and VBP when the AV begins to cut back into the running lane.
    
    \item \emph{Cut-in Angle} ($\theta$) is the steering angle taken during the cut-in move.
\end{itemize}

Taken as an ensemble, the last four parameters in the above list define a \emph{driving profile} for the manoeuvre, which characterises a level of urgency or aggression associated with the manoeuvre. An `aggressive' profile might be taken as one which has a low pull-out clearance, high pull-out angle, low cut-in clearance and high cut-in angle, indicating a `sharp and close' manoeuvre taken by the AV around the VBP. A more `relaxed' driving profile may have larger clearances and lower steering angles. Various profiles of manoeuvre with different characteristic values are assessed in the experimental work discussed later. It should be noted that other UKHC rules place (at least qualitative) restrictions on these parameters to ensure safe driving behaviour (see the fourth bullet-point of Rule 163.

As with the problem of direct translation of the UKHC rules, the approach aimed to develop a model using the concepts of the natural language of the rule rather than by other parameters, for example time-to-collision, which has been found in human factors studies to be the most significant parameter characterising overtaking manoeuvres \cite{lenard2018, Chen2015}. Hence the model developed for the Safe Distance Ahead measurement differs from many that have been developed for analysis of overtaking manoeuvres, which aim to capture the decision making of the driver rather than the characteristics of the legal code of behaviour that is essentially an independent observer's perspective on the scenario.

\section{Assertion-checking System Application}

\subsection{Database Design}

\begin{figure}
    \centering
    \includegraphics[width=8.5cm]{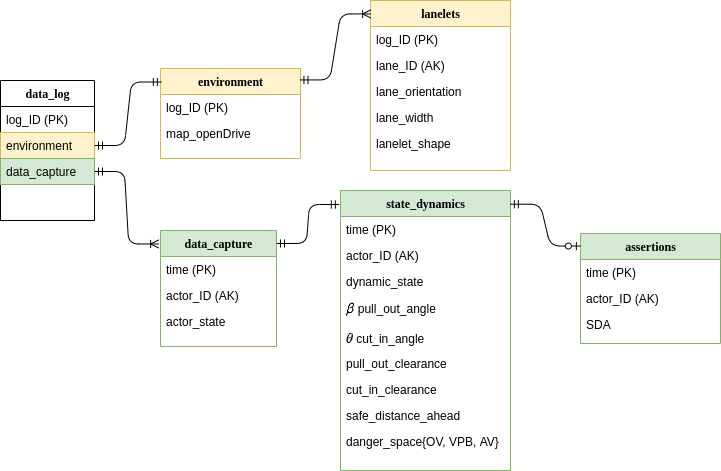}
    \caption{PostgreSQL database Entity Relationship Diagram.}
    \label{fig:erd}
\end{figure}

Analysis of the overtaking manoeuvre problem fits neatly into the geospatial database framework for assertion checking. The vehicles can be represented as shape objects using the PostGIS system extension which allows for logical tests and measurement metrics to be efficiently calculated between shapes. An Entity Relation Diagram (ERD), Figure~\ref{fig:erd}, shows the variables within the database and the relationships between them\footnote{See \url{https://en.wikipedia.org/wiki/Entity-relationship\_model} for an explanation of \textit{crow's foot} notation.}. 

For each log file there exists an environment table containing map information, which in this case is based on the OpenDrive\footnote{\url{https://www.asam.net/standards/detail/opendrive/}} and lanelet formats. From the lanelet data we derive a table containing useful entries such as lane width and orientation for the environment. Data captured from the simulation (or runtime monitoring), contains a state vector for the actors in $data\_capture.actor\_state$ which contains position and orientation information. From this captured data table we can calculate dynamic state information, that is information changing over time (e.g.\ velocity, acceleration) or with vehicle speed (e.g.\ braking distance). 

Further insight to the driving situation can be derived when combining actor state information with information about the environment. For example, driving lane orientation ($lanelets.lane\_orientation$) can be used to calculate the vehicle heading angle relative to the driving lane, to infer the cut-in or pull-out angles ($state\_dynamics.pull\_out\_angle$) required for the Safe Distance Ahead calculation. The final $assertions$ table is where the results of the assertion can be written and stored. The assertion compares the observed distance ahead with the required Safe Distance Ahead, and indicates whether a situation is safe as a Boolean pass/fail result where a `pass' result is generated if the observed DA is greater than the SDA calculated from the driving profile being followed by the AV.

\subsection{Modes of Use of the Assertion-checking Subsystem} \label{Assertions_at_Sim_Time} \label{assertions_at_run_time}

Assertion-checking systems can be used 
in a variety of ways, both during simulation and in runtime monitoring:


\underline{\textbf{Modes of use during simulation:}}
	
	\begin{itemize}
		\item \emph{Simulation Control (Halting)}: Assertions can be checked as the simulation runs, and a simulation could be halted after a major safety violation if there is little value in continuing with a simulation beyond that point.
		
		\item \emph{Operator Warnings}: 
		As above, assertions can be checked as the simulation runs but, instead of halting, the operator can be warned of the occurrence. Thus the operator can exercise their judgment on whether to continue or not. 
		We anticipate that this may be the best policy for minor safety violations or events such as collisions which may lead to secondary effects like non-deterministic behaviour in the simulator~\cite{chance2021}, where the log file data might be degraded or unrepeatable following the event. 
%
		\item \emph{Simulation Log Annotation}: for negligible safety assertions failures, or for non-safety checks such as service performance monitoring, assertions could be checked as the simulation runs. However, instead of halting or providing warnings, assertion failures could simply result in annotations being added to the simulation log data, indicating that an assertion failed (or also possibly for near-miss cases) at a given time step but otherwise having no other externally observable effect. The log can then be analysed retrospectively after the simulation run has finished.
		
		\item \emph{Retrospective analysis}: Instead of checking assertions as the simulation runs, the captured data can be recorded and passed to the assertion checking database after simulation has finished, to analyse an AV's performance retrospectively. 
	\end{itemize}

	The simulation experiments with assertions presented in Section~\ref{Experimental_scenario} were all run retrospectively, as they serve as demonstration of concept rather than full implementation of all the above modes of use. 


\underline{\textbf{Modes of use during runtime monitoring:}}
	
	\begin{itemize}
		\item \emph{Warnings and Alarms}: Assertion checks can be used to raise warnings or alarms to human vehicle operators, to draw their attention or request them to take control over the vehicle if safety violations occur. Additionally these warnings can be used to indirectly assess the performance of subsystems during runtime to provide development feedback in future versions of the system.
		
		\item \emph{Triggering safety functions}: Detection of assertion violations can be used to initiate automated pre-planned risk mitigation plans as a safety function, where such occurrences present an immediate and sufficiently high risk that such a reaction is warranted~\cite{RuntimeReview}. Having a ground truth reference to judge on monitored data at runtime can be challenging. Assertion failures can provide a reliable verdict to execute appropriate risk mitigation plans.
		
		\item \emph{Self-adaptation}: using assertion checking during runtime can be combined with measurements of key and safety performance indicators (KPIs, SPIs)~\cite{Koopman2020} or other scoring functions to help the AV 
		to actively learn and reliably adapt their behaviour during runtime~\cite{Kang2020}.
		
		\item \emph{Accident investigation or insurance risk assessment}: Assertion-checking subsystems can be used to annotate the internal data logs of an AV to help identify potential causes of an accident in any post-incident investigation (similar to flight data recorders in aircraft), or to measure the occurrence rate of UKHC driving rule infractions for insurance risk adjustment purposes. However, there are ethical concerns associated with such applications, as discussed in~\cite{EthicalBlackBox}.
		
	\end{itemize}

\section{Experimental Case Studies} \label{Experimental_scenario}

We developed early worked examples of assertions and measurements, and applied these to both simulation data and real-time video footage.
%
%
%
While these examples initially served as an aid to shaping the methods and techniques described previously, we use them here to illustrate the general concept and utility of the proposed approach.
The use of methods such as modelling, and the need for features such as the use of standard procedures/library functions, all emerged as direct consequences of developing these examples.

In the following we present two example case studies inspired from an overtaking scenario; the first aims at demonstrating the use of assertions in simulation testing, and the second shows how assertions can be used as \emph{safety performance indicators} (SPIs) to measure how well an AV or other road user maintains the relevant safety properties at runtime.

\subsection{Experimental Scenario} 
\label{carla_challenge}

To demonstrate the functionality of the assertion based safety validation process, we consider an overtaking scenario similar to Scenario 6 of the driving situations in the NHTSA\footnote{\url{https://www.nhtsa.gov/}} pre-crash typology~\cite{nhtsa_precrash} that were published in the CARLA Challenge.\footnote{\url{https://carlachallenge.org/}} The corresponding UKHC rules~\cite{highwayCode} that apply to this scenario are Rules 162-169. Specifically, the scenario involves overtaking on a single-lane road where the test vehicle must perform a lane change to avoid a parked or broken down vehicle. 
The vehicle must leave a sufficient gap to the vehicle it is overtaking and also not endanger any vehicles that are oncoming.



In the simulation and runtime monitoring experiments, we chose to assess clauses in two different UKHC overtaking rules. The simulation study developed an assertion for the first bullet-item of Rule 162 (as discussed in Section~\ref{interp_nat_lang_UKHC_rules}). 
The assertion checks whether a Safe Distance Ahead exists for an AV performing an overtake manoeuvre.
The runtime safety validation study developed video analysis measurements that would support assertion checks against the fourth bullet-item of Rule 163 
(see the italicised clause in Section \ref{interp_nat_lang_UKHC_rules}).
While the measurements were not developed fully into assertions in SQL code, the study does capture the information essential to the assertion hypothesis for that UKHC clause (the Cut-in Clearance distance as discussed in Section~\ref{model_based_analysis}), which can be generated as a derived variable in the database using a standard procedure function as discussed previously.

\subsection{Simulation Case Study}  \label{sim_case_study}
We developed an assertion check for Safe Distance Ahead at the start of overtaking on a single-lane road, just as the vehicle begins to cross the road centre line. The assertion uses the expression for SDA discussed earlier in Section~\ref{model_based_analysis}. 

\subsubsection{Environment model}
The simulation used the map segment shown in Figure~\ref{fig:assertion_database_annotated}, which depicts a road in Bristol, UK, that is part of an AV test route being considered for the ROBOPILOT project.\footnote{\url{https://www.robopilot.co.uk/}} 
For the sake of simplicity the simulation was executed on one of the straight sections of road, which avoided having to take considerations such as obscuring objects or limited lines of sight into account. The road layout model was developed using the Road Runner software package, from an OpenDrive environmental model developed from Open Street Map (OSM) data obtainable freely from UK government sources (Ordnance Survey).\footnote{\url{https://os.openstreetmap.org/}} 



As discussed in Section~\ref{geospatial_database}, lane polygons are required for efficient measurement of vehicle lateral movement within, and partial occupancy of, road lanes. The \textit{lanelets}~\cite{lanelets2014} lane partitioning concept allows exactly this level of description of the scene. 
We used an open-source software conversion package~\cite{lanelets_to_openDrive} to generate the lanelet data from the OpenDrive~\cite{opendrive} model, dividing the roads into individual lanes and lanelets. This data was imported into the PostgreSQL database as a table of lane segment polygon objects ($lanelets.lanelet\_shape$ in \ref{fig:erd}) covering the whole road layout used for the study. Additionally, information about lane boundaries and orientations, required to identify road orientation and running lane directions, was appended to this data. 
Figure~\ref{fig:assertion_database_annotated} shows an example of the lanelet shape object overlaid onto a road network. The blue boxes represent the dynamic actors, which in our case are the vehicles involved in the overtaking scenario. 
All of the above world map information was pre-loaded statically into the database, for use by the SQL query engine running the assertion checks.

\subsubsection{Overtaking profiles} \label{overtaking_profiles}

The intent of the Safe Distance Ahead assertion we developed was to assess whether sufficient distance ahead of the simulated AV was available at the initial decision point of the manoeuvre for it to complete an overtaking manoeuvre safely. However, in practice, sufficiency of the distance ahead is dependent in part on the driving profile that the AV controller has selected to perform (see Section \ref{model_based_analysis}). But, if the safety validation exercise is to be performed as an independent `black-box' measurement of AV behaviour (as is often the case in many practical development projects), the internal plans of the AV may be unavailable to the system testers, and we conducted our experiment in such a manner as to reflect such a situation.


\begin{table}[t]
\centering
\begin{tabular}{lcccc}
\hline
\\
\multirow{3}{*}{\textbf{Scenario}} & \textbf{Achieved} & \multicolumn{3}{c}{
\textbf{Required SDA by profile (m)}}\\ 
& \textbf{Distance} & `Relaxed' & `Nominal' & `Aggressive' \\
& \textbf{Ahead (m)} & 101.39 & 63.73 & 40.02 \\
\\
Safe & 76.43 & FAIL & PASS & PASS\\
Near miss & 58.33 & FAIL & FAIL & PASS\\
Collision & 35.63 & FAIL & FAIL & FAIL\\
\\
\hline
\end{tabular}
\caption{Pass/Fail Assertion results.} \label{Overtaking_Profiles}
\end{table}

In the experiment the AV was programmed to follow a preset trajectory. In each simulation run the starting location of the oncoming vehicle was changed, such that at the start of the overtake manoeuvre the distance from the OV to the AV was different. For each of these three simulation runs the DA achieved in the simulation was compared with the required SDA for each of three different driving styles. Where the achieved DA was less than or equal to the required DA of a driving style the assertion check marked the comparison as a failure (which indicates that the situation was unsafe). These results are shown in Table~\ref{Overtaking_Profiles}.


We defined three different driving profiles (refer to \ref{model_based_analysis} a description of the parameters) corresponding to three different levels of urgency of overtake manoeuvre (called `relaxed', `nominal' and `aggressive'). %
The levels of urgency were based loosely on similar concepts developed in other Human Factors research~\cite{Chen2015,wang2019analysis}, which had equated the level or degree of urgency with \emph{Time To Collision} (TTC). These sources identified that a TTC of approximately two seconds corresponds approximately to a medium degree of urgency, which we have labelled `nominal'. 

\subsubsection{Simulator system}

The setup used for the case study followed the assertion monitoring architecture described in Section~\ref{generic_sim_system}. 
The system used the CARLA simulation software~\cite{CarlaSimulator}, an open-source autonomous driving simulation environment. Dynamic state data generated by CARLA for the simulated actors was passed to a PostgreSQL database as shown in Figure~\ref{fig:SimulatorArchitecture}, with the SQL assertion check being applied in retrospective analysis mode as discussed in Section~\ref{Assertions_at_Sim_Time}.
%



\subsubsection{Simulation experiments}

\begin{figure}
    \centering
    \includegraphics[width=8.5cm]{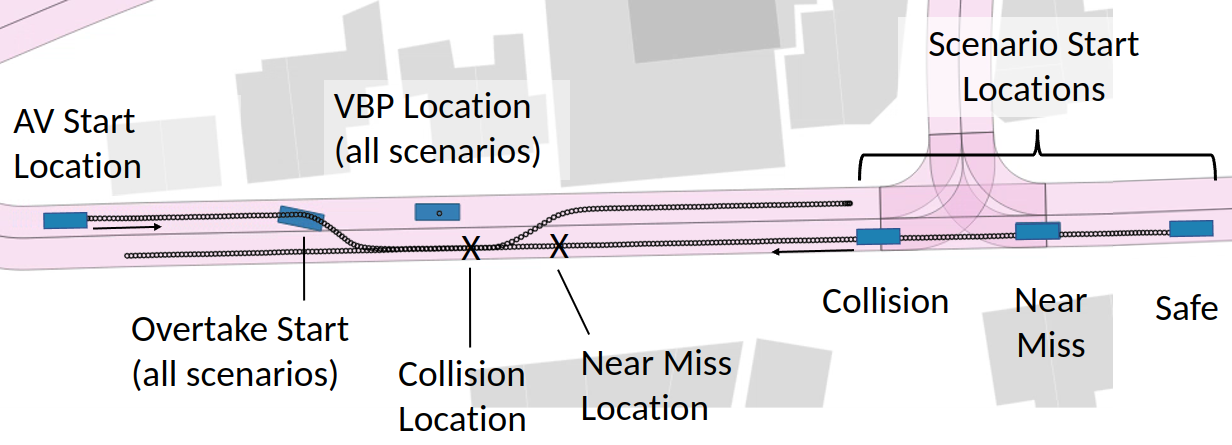}
    \caption{Simulation experiment driving paths for each scenario.}
    \label{fig:Sim_experiment_layout}
\end{figure}

Three simulation runs were performed, in which the simulated AV followed the same trajectory, overtaking a stationary (parked) VBP as shown in Figure \ref{fig:Sim_experiment_layout} at a constant simulated speed of 25mph along a section of single carriageway approximately 150m long. In each simulation run, the OV followed a straight trajectory down the oncoming lane at a constant speed of 25mph, but in the start location of the OV was varied so that as the AV began its overtake manoeuvre, its separation from the OV (the DA) varied. In the run labelled ``Safe'', there was clearly sufficient distance for a safe overtake; the run labelled ``Near Miss'' produced a situation where the AV and OV came very close to colliding (within 1m distance), and in the third run (labelled ``Collision'') a head-on collision occurred. The locations  relative to the VBP of the near miss and collision events are marked on Figure \ref{fig:Sim_experiment_layout}.
Using fixed vehicle paths facilitates evaluation of the assertion, as it makes the test repeatable within a deterministic simulation environment~\cite{chance2021}. 

%



The three simulation runs described previously produced simulation data logs that were subjected to assertion checking to ensure that a Safe Distance Ahead existed as the AV began to cross the road centre line into the oncoming lane at the start of the overtake manoeuvre. However, there are several factors conditioning the test results that need to be taken into account. 
Comparison of the three variations of test against the simulated scenarios (safe, near miss and collision) produces nine distinct tests. 

The fixed overtaking manoeuvre trajectory that had been pre-set for the AV vehicle approximated to a medium level of urgency, in terms of the pull-out and cut-in clearance distances achieved (about 2s time to collision at the simulated speeds of the vehicles). This meant that for a set of tests having a range of urgency levels centred on the medium urgency case, one should get a sliding scale of test case passes and failures as the test case variants are applied to decreasing initial separation distances between AV and OV in the three simulation logs.


\subsubsection{Simulation-based assertion checking results} \label{sim_results}

The results of all tests are provided in Table~\ref{Overtaking_Profiles}, which shows the required separation distances at the start of each manoeuvre (determined by the modelling analysis formula described in Section~\ref{model_based_analysis}), the Distance Ahead that was achieved in each simulation run, and hence the pass/fail assertion results of comparing the two distances in each case.

The table shows that it would have been inadvisable to perform a `relaxed' overtake in any of the three scenarios presented, as this would have resulted in a near miss or a collision. If a `nominal' overtake is attempted (which actually corresponds to the behaviour of the AV in the scenarios), this policy was acceptable for the `Safe' scenario, but inadvisable for the `Near Miss' and `Collision' scenarios (as their descriptive names imply). Finally, if the AV had been programmed to overtake aggressively in these scenarios, then only in the case of the Collision scenario circumstances would the Distance Ahead have been short enough for the manoeuvre to be inadvisable; in the other two cases the overtake could have been achieved successfully (although aggressive overtaking may result in other safety assertions being violated).

\subsection{Runtime Monitoring Case Study}
In our experimental study of runtime monitoring we developed an assertion monitor for checking safe stopping distance during the overtake. However, the SDA check is only applicable at the \textit{start} of an overtake - it says nothing about whether separation distances \textit{remain} safe at later stages of the manoeuvre.

In many real driving scenarios, environmental conditions that may have been checked and found safe at the beginning may cease to be true or valid as the scenario progresses, and driving plans or actions may need to be altered midway through as a consequence. Therefore, it can be highly advisable for autonomous vehicles to evaluate assertions continually during runtime, to detect the onset of hazards as a cue for changes of strategy.
To demonstrate the use of assertions at runtime, in the absence of a roadworthy AV controller, we analysed public domain video footage of an attempted overtaking manoeuvre in a manually driven car.\footnote{Link to video used in analysis: \url{https://youtu.be/Gzi4X7WZkRI} time period 01:30 to 01:45 minutes.} The video shows a scenario within a road journey, where an overtake manoeuvre was initiated because the road ahead appeared clear in the opposite lane to allow a safe manoeuvre. However, during the manoeuvre an oncoming vehicle, which had been obscured by a combination of a slight dip and a minor bend in the road, suddenly became visible at a distance that did not allow the manoeuvre to be completed safely, and the human driver was forced to abort the overtake and pull back into the running lane behind the VBP \footnote{The cause of this error was a mistake in estimating the safe distance ahead, which should have treated the observable line of sight as the available distance ahead, in the absence of any directly visible OV. Since we were only intending to demonstrate the principle of assertion checking in our simulation study, we eliminated this complication from the simulation scenario deliberately, by using a map of a long straight road section where an OV would always be visible.}. Such situations occur frequently in typical (complex) real-world road layouts, if vehicles change speed or if for any other reason the SDA calculation at the start of the manoeuvre gives misleading estimates. 

The video example shows the advantage (perhaps even the necessity) of continual re-evaluation of safety properties during runtime. If an AV had been fully committed to such an overtake manoeuvre after making only one assertion check at the start, a collision might have occurred.
%

%

\subsubsection{Definition of assertions}
\label{DS_assessment}
In the video analysis study, the aim of the assertion checking applied to runtime data was to assess whether the AV risks colliding with any of the other vehicles involved in the scenario during the overtake and in turn take an informative decision on aborting or continuing the overtake. The assertions defined for this study were therefore related to the danger spaces (DS) of the AV and OV, as defined by the UKHC.

%
Following Rule 126 of the UKHC, the basic requirements being evaluated during the different stages of the overtake (pulling out, passing, and cutting in, as described in Section \ref{model_based_analysis}) were:
%
%
%


\begin{enumerate}[i)]
\item \emph{Safe separation with VBP:}
The AV should maintain safe separation with the VBP throughout the overtake manoeuvre; in practice this is primarily applicable during the Pulling Out phase.
\begin{itemize}
	\item Assertion condition: VBP outside DS$_{AV}$
\end{itemize}
\item \emph{Hazardous condition (requiring urgent avoiding action):} 
As two vehicles moving in opposing directions in the same lane converge, several conditions associated with their relative proximity develop, which can be checked by assertion and used as cues for safety related decision making such as aborting manoeuvres or taking avoiding actions. Any situation, which requires such action to be taken is called a \textit{hazardous situation}. An overtake manoeuvre situation will begin to become potentially hazardous as the danger spaces of the AV and OV begin to overlap. At this stage a collision might occur if neither vehicle takes any avoiding action, so this overlap condition measures the onset of the hazardous situation.
\begin{itemize}
	\item Assertion condition: no intersection between DS$_{AV}$ and DS$_{OV}$
\end{itemize}
\item \emph{Imminent collision conditions:}
As a potentially hazardous overtaking situation develops, other conditions related to the direct occupancy of one vehicle inside another danger space indicate the increasing imminence of a collision, and can be used as cues to reinforce further the need to take avoiding action:
\begin{itemize}
	\item Assertion condition: OV outside DS$_{AV}$ 
	\item Assertion condition: AV outside DS$_{OV}$
\end{itemize}
\end{enumerate}
These four assertions were evaluated in the selected video file by means of video processing analysis.

\subsubsection{Video processing}
%
We used the following approach to approximate the processing of the real-time data stream for the runtime case study. 
First, a pre-trained object classifier, YOLOv3~\cite{Yolo}, was used to provide locations of vehicles in each frame and how much area each vehicle occupies of the frame. 
%
To determine separation distances between the AV and other vehicles 
the longitudinal and lateral separation distances were calculated between other vehicles and the AV.
The longitudinal distance ($s$) of a vehicle along the road from the AV can be estimated using Equation~\ref{s_equation}:
\begin{equation}\label{s_equation}
    s = \frac{c\times W}{w}
\end{equation}
where $c$ is a constant capturing the focal length of the camera and is found empirically, $W$ is the actual width of a vehicle in meters, and $w$ is the width of the box detected by YOLOv3 in pixels. 
Second, the lateral distance $d$ of the AV from the middle white line is estimated using Equation~\ref{d_equation}: 
\begin{equation}\label{d_equation}
    d = \frac{w_{L (real)}}{w_{L (pixels)}} \times d_{pixels}
\end{equation}
where $w_{L (real)}$ is the real lane width measured in meters, $w_{L (pixels)}$ is the lane width measured in pixels and $d_{pixels}$ is the lateral distance of the AV from the middle white line in pixels. $d_{pixels}$ is determined by simply finding where the middle white line is when compared to the middle of the frame. The VBP and OV are assumed to have a fixed lateral distance from the white line separating lanes. 
Third, the pull out $\beta$ and cut in $\theta$ angles of the AV are calculated using the change in lateral distance and change in the longitudinal distance measured. Fourth, as there is no vehicle speed associated with the video, we based the DS calculation on the worst case scenario; that is all vehicles were travelling at the speed limit of the road. 
According to Rule 124 in the UKHC the national speed limit of the single carriage way in the video for cars (AV and OV) is 60mph, and 50mph for goods vehicles (VBP). Using the above steps the real time video was converted to a 2D driving scenario, where the Danger Space assertions listed in section~\ref{DS_assessment} can be checked.

\subsubsection{Runtime validation results}
\begin{table}
	\centering
	\begin{tabular}{ clcc }
		\hline
		\\
		\textbf{Assertion} & \multirow{3}{*}{\textbf{}} & \multicolumn{2}{c}{\textbf{Overtaking stages}}\\
		\textbf{ No.} & \textbf{Assertion} & \textbf{Pulling out} & \textbf{Passing VBP} 
		 \\
		&  &  &   \\
		(1) & VBP outside DS$_{AV}$ & PASS & PASS  \\
		(2) & OV outside DS$_{AV}$ & PASS & FAIL  \\
		(3) & AV outside DS$_{OV}$  & PASS & FAIL  \\
		(4) & No DS$_{AV}$ $\cap$ DS$_{OV}$  & PASS & FAIL  \\
		%
		%
		%
		\\
		\hline
	\end{tabular}
	\caption{Pass/Fail danger space assertion checks results for the runtime monitoring study. Note: The use of intersection symbols in the table is shorthand for a `does not overlap' relation between polygons.} \label{Overtaking_DS_table}
\end{table}

\begin{figure}
    \centering
    \includegraphics[width=8.5cm]{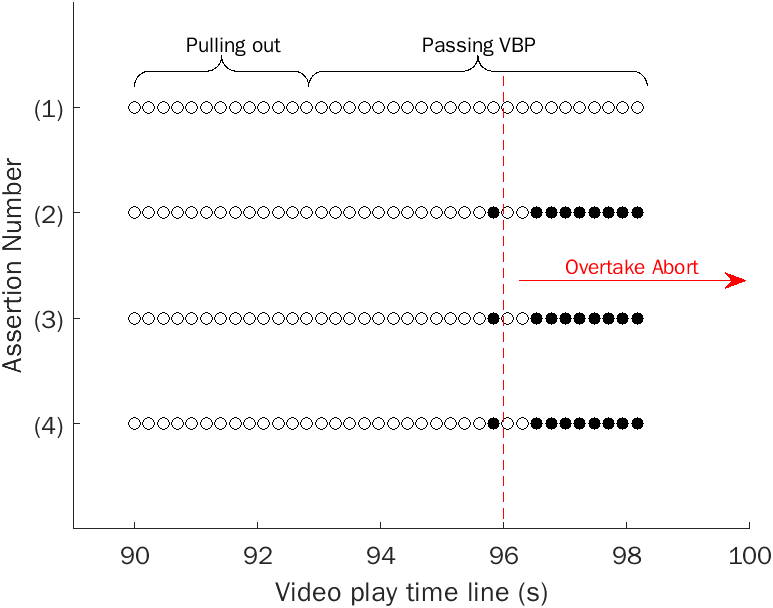}
    \caption{Assertion checking results for runtime case study.}
    \label{fig:Runtime_timeline}
\end{figure}

Table~\ref{Overtaking_DS_table} shows the results for the four runtime assertions introduced in Section~\ref{DS_assessment}. 
%
The four assertions were checked against every frame, and were found to change from ``PASS'' to ``FAIL'' approximately six seconds after the start of the overtake manoeuvre, as the oncoming vehicle suddenly became visible. 
Due to the degradation of video quality, caused by prevailing ambient conditions (light rain), and the small size of the OV object image in the video frame at that time, a transient state change occurs as shown in Figure~\ref{fig:Runtime_timeline} at 96s, where the evaluation of the assertions reverts back to ``PASS'' for a few frames before going back unambiguously to ``FAIL''. 
This emphasizes the need for good quality environmental sensing (e.g. high resolution cameras for more reliable object classification) to obtain clearly defined changes of state.
In the absence of good quality sensing due to harsh environmental conditions, autonomous systems can benefit from the use of classifiers that adapt to operational environments~\cite{ghobrial2022operational}, giving more reliable performance in harsh environmental conditions that were not trained on at design time. 
Figure \ref{fig:Runtime_timeline} also shows that all safety assertions fail simultaneously. This is due to the fact that the obscuration of the OV was cleared with only a short separation distance between the ego-vehicle and the OV, so assertions (2)-(4) all failed immediately as soon as they became measurable. In a scenario where the OV is visible from the start, one would expect assertion (4) (no overlap between DS$_{AV}$ and DS$_{OV}$) to fail first, followed shortly by (2) and (3) as the vehicles drew closer together.

The change of evaluation state of the assertions part way through the overtake manoeuvre clearly shows the value of applying them as runtime monitoring checks. For example, the transitions can be used as input signals to the AV control system to prompt an emergency avoidance action.

\section{Discussion} \label{discussion}

The initial work presented in this paper has raised a number of consequent issues that require further investigation.

\subsection{Assertion Checking to Measure Performance}

While we have applied the assertion checking methodology to measure the safety properties of vehicles in simulation and at runtime, the methodology can also be extended to validate other important system properties. This may include performance based metrics such as smooth driving (minimal jerk), accuracy of control (such as parking or stopping precisely, e.g.\ at a bus stop), and decisive behaviour at junctions or roundabouts, sometimes termed \emph{liveness}~\cite{kim2014mpc}. In particular, the mission performance of a vehicle, for example, can be measured in journey time or efficiency of movement through traffic, which may affect not only passenger expectations but also the commercial viability of such an autonomous system. 

Performance (achievement of task goals) and safety (avoidance of task hazards) are distinct behavioural properties of a system; achievement of one does not necessarily entail the other. A task controller must achieve satisfactory compliance with both properties if a dependable system is to be developed. In some scenarios, such as during overtaking, there may exist a competition within the controller to satisfy both the achievement of goals and the avoidance of hazards. 

For example, an autonomous taxi may need to perform a minimum number of fare-paying journeys in a day in order to remain commercially viable. Performance assertions can measure whether suitable opportunities for manoeuvres such as overtaking were taken efficiently, without incurring delays or driving sub-optimally for the given road network i.e. using all available space. However, it must still drive safely as it conducts each journey, so being unduly aggressive in its driving may cause other risks in terms of other interactions with traffic. Assertions can measure how the AV trades off these two properties in the execution of its task(s). 

\begin{figure}
    \centering
    \includegraphics[width=8.5cm]{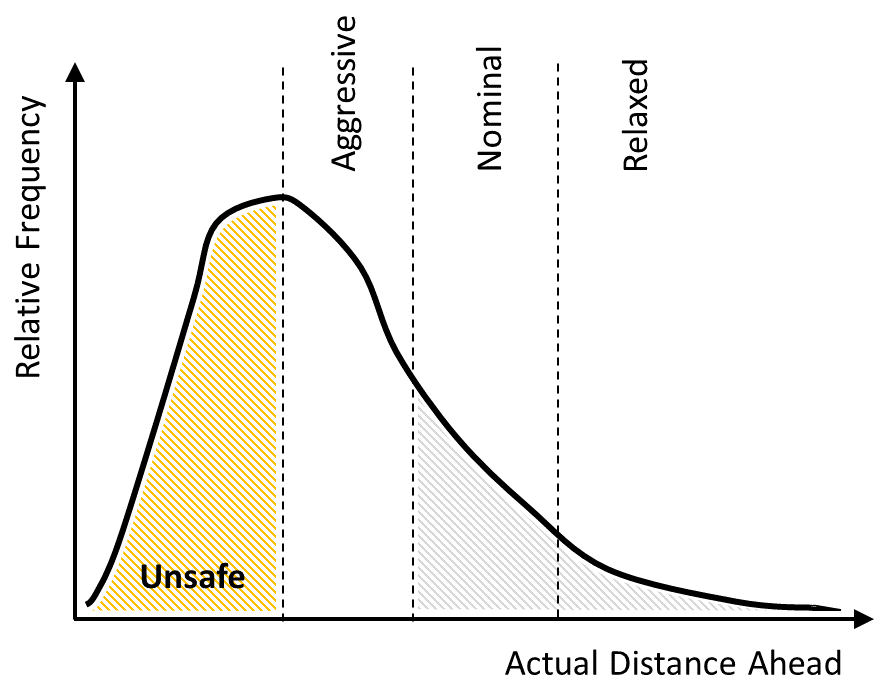}
    \caption{Assumed distribution of overtaking opportunities as a function of actual distance ahead.}
    \label{fig:distribution}
\end{figure}

An analysis for overtaking performance can be observed in the simulation results presented in Section~\ref{sim_case_study} by comparing the required SDA to the actual distance ahead of the Oncoming Vehicle and the relative frequency with which overtaking opportunities of a given distance ahead actually occur. Figure~\ref{fig:distribution} presents a sample of overtaking distances based on findings presented by Farah~\cite{farah2016drivers} as a representative example. Some opportunities may be unsafe because the actual DA is less than the required SDA (as illustrated in region of the graph marked `unsafe'). For a given driving profile the proportion of those overtaking opportunities that are achievable are those where the actual DA is greater than distance that can be achieved under that profile. Taking the nominal profile as an example, this is represented in Figure~\ref{fig:distribution} by the shaded area under the curve to the right of the nominal profile line. 

From the point of view of performance, the AV should make best use of the available gaps to overtake and may need to change its driving profile in order to do so. 
However, aggressive driving profiles can introduce new risks, e.g. adverse reaction from other drivers and the UKHC requires drivers to maintain the minimum level of aggression necessary for overtaking.
If a vehicle chooses a driving profile that is overly conservative, it will reject opportunities to overtake that would be safe if a more aggressive profile were selected, and hence its performance would suffer. 

We assume that the frequency at which overtaking opportunities of a given actual distance ahead may occur as illustrated in Figure~\ref{fig:distribution} (based on findings from~\cite{farah2016drivers}). 
In general the more aggressive the driving style selected, the higher the number of opportunities that can be taken. 

\begin{figure}[t]
    \centering
    \includegraphics[width=8.5cm]{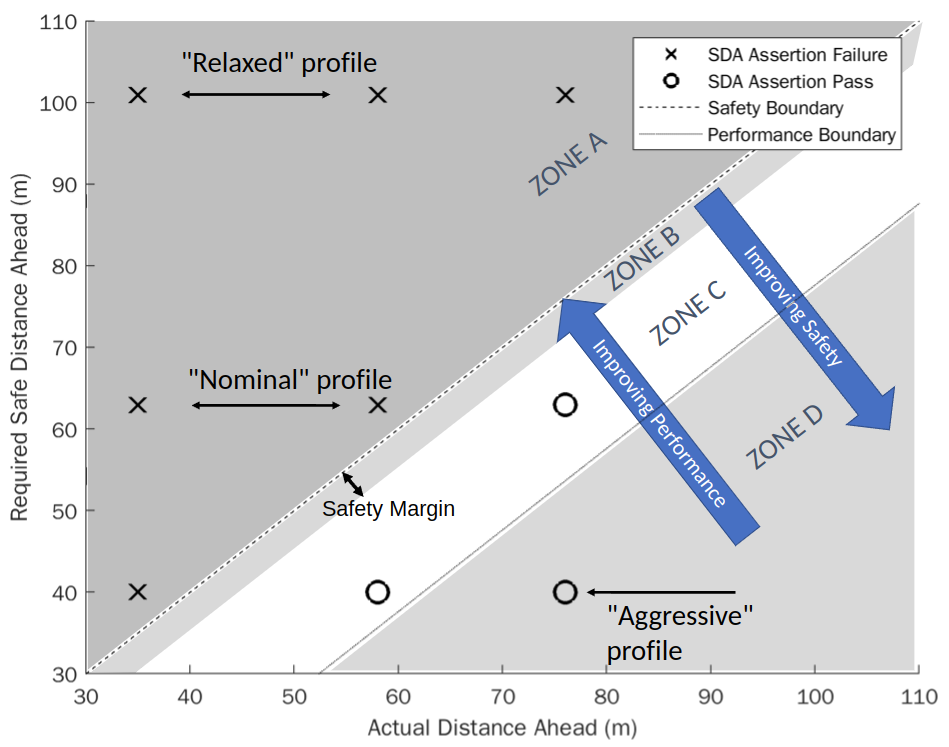}
    \caption{Simulation results for performance.}
    \label{fig:performance_results}
\end{figure}

For the results of our simulation experiment the trade off between driving style, DA and SDA is shown by the zones in Figure~\ref{fig:performance_results}. 
Zone A includes all situations where the actual distance ahead is less than the SDA and are therefore unsafe. Zone B is a required safety margin that is used to  excludes all near-miss situations which is typical practice in many safety related applications. 
Zones C and D describe situations where the ratio of DA/SDA is  getting progressively larger and represent increasing levels of safety. 
Zone D, which captures all situations where there is more than 2.5s TTC between the AV and the OV during the overtake, represents situations where there is so much safety that a decision not to overtake would be overly conservative. Zone C captures all situations where more overtaking opportunities are taken but safety is still maintained. 
A performance assertion should measure whether the AV is in Zone C which is a trade off between driving profile, the actual distance ahead and the required safe distance ahead. The optimal overtaking condition is therefore the least aggressive driving profile that attains Zone C for any given ratio of DA/SDA. 
Switching from a conservative to a more aggressive profile may allow an AV in Zone A to move into Zone C and take opportunities to overtake that would be denied if the driving profile is fixed. 

There is therefore a trade-off between safety and performance for optimal driving behaviour, where performance is improved by moving towards Zone B (but not passing the safety margin), and safety is improved by moving towards Zone D (see large arrows in Figure~\ref{fig:performance_results}). While we have observed this conceptual trade-off between safety and performance in our analysis, we also see evidence of this in the UKHC rules themselves. An example of this is the 4th bullet point in Rule 163 (see Section~\ref{interp_nat_lang_UKHC_rules}) where the overtaking rule states ``Move quickly past the vehicle'' and ``Move back to the left as soon as you can...'', which suggests that performance can play a part in safe driving behaviour. 
We see these competing demands resolve in optimal driving behaviour analogous to that described by Zone C in Figure~\ref{fig:performance_results}. This analysis considers only the overtaking scenario and in reality there will be many competing demands on the decision making process of a controller.

\subsection{Assertion Checking of Rules about AV Intentions, Planning and Reasoning}

One significant topic of future study relates to the extent to which assertions require access to states of internal decision making within the autonomous vehicle being monitored. The taxonomy of assertions presented in Section \ref{assertion_categories}, and the methodology we have developed for deriving them from UKHC rules, is applicable to rules that define externally observable states or behaviours of the AV in its road environment. However, many UKHC rules are advisories to drivers 
regarding potentially hazardous situations, making recommendations on how to actively perceive, deliberate, and select special or particular action to avoid harmful events. These rules (or clauses within rules) cannot be validated purely by external measurements of situated state and behaviour alone - they require information about the internal state of the AV algorithms, to determine whether those algorithms have been making perceptions, plans or action selection decisions as these kinds of rule recommend.

For example, referring to UKHC Rule 163 listed in Section \ref{interp_nat_lang_UKHC_rules}, the second bullet-item provides the following advice about the sequence of perceptions and decisions related to overtaking: \textit{``use your mirrors, signal when it is safe to do so, take a quick sideways glance if necessary into the blind spot area and then start to move out''}. The rule requires that drivers make specific perceptions, for example into the vehicle blind spot area, prior to any decision to change lane. However, this rule clause can only be validated by checking the internal perception and reasoning of the AV control system to confirm that it had made such perceptions and decisions in the preferred order. This cannot be inferred by measuring the (externally objective) position or velocity of the vehicle; instead, the AV control system must declare what internal perception or reasoning it may have made in order to select a given action.
This example highlights the need for \emph{explainable AI}, a topic gaining increasing attention in the AI, robotics and autonomous systems community. The opinion is widely held \cite{deeks2019judicial, o2019legal, wortham2020transparency} that it may be essential for any such system employed in a safety related application to be able to explain its decision making.
In many AI technologies the internal representations of perceptions or decisions are complexity and opaque, which makes the task of identifying, by manual inspection, how a system might have come to a particular decision extremely difficult. When combined with the non-deterministic nature of complex operational situations in real world environments, this means that particular decisions may not be repeatable in post-incident analysis without the system being able to report the precise situation that it perceived, and the planning or action-selection decisions that it made as a consequence. 
%
This creates a significant opportunity for the use of assertion-based runtime monitoring as discussed in Section~\ref{assertions_at_run_time}.

The existence of 'cognitive rules' in the driving code of conduct suggests that the taxonomy of assertion types described in Section \ref{assertion_categories}, while a comprehensive classification of assertions for 'objective rules', may require extension to address the classification of assertions for cognitive rules. We will address this issue in future work.

\section{Conclusion}\label{conclusion}

In this paper we have presented a methodology to systematically derive formal logical expressions, encoded as assertions, from the UKHC. We have introduced a process for monitoring assertions that can be used for simulation-based assertion checking and runtime monitoring during operation.
We have also developed a taxonomic classification of assertion types that are applicable to checking of rules that can be evaluated by objective measurement. This scheme reveals key aspects of assertion design, such as assertion reference points and trigger expressions. These should be incorporated in any program code to execute the checks to ensure that assertions are expressed correctly. We have developed SQL-based programming schemes for encoding the assertions, and demonstrated their use on a PostgreSQL/PostGIS database, which receives simulation trace data from a Carla simulator running an AV simulation scenario. We have also applied assertion checks to sample video data to emulate the use of assertion checking as a runtime monitoring technique.
The results obtained from both the simulation and the runtime video analysis case studies show the general feasibility and utility of assertion-based safety validation, with the assertions being derived from regulatory documentation written in natural language and intended for human drivers rather than written specifically to be usable by computer. The results provide our first indication that human-oriented natural language rules for driving can be an effective 
basis
for 
safety validation of autonomous vehicles.
We also present how the UKHC-based assertion checking methodology can be used as a method to infer autonomous vehicle performance and 
is equally as usable in runtime monitoring as it is in simulation.

We are considering numerous directions in which to continue this work such as extending the sources of information for safe driving, e.g.\ using the German driving regulations~\cite{acountability}, and models such as the RSS model~\cite{RSS_Shalev_Shwartz2017,RSS2_Koopman2019}. We are also researching the use of the testbench to perform falsification testing~\cite{corso2020survey, akazaki2017causality} and \emph{situation coverage testing}~\cite{alexander2015} and how other techniques such as \emph{Environmental Survey Hazard Analysis}~\cite{harper2021towards} can be integrated with the methodology. We also aim to investigate assertion checking to provide reward function scores as a critic in reinforcement learning for continuous improvement of the AVs balance between safety and liveness during operation.

\balance

\printbibliography

\newpage
\begingroup
\let\clearpage\relax 
\onecolumn 
\section*{Appendix A}\label{appendix_a}
Figure \ref{fig:overtaking_analysis_diag} presents a model of an overtaking manoeuvre, showing the derivation of a formula for calculating Safe Distance Ahead, which was used in the simulation study results presented in Section \ref{sim_case_study}.
\begin{figure*}[h]
    \includegraphics[width=18cm]{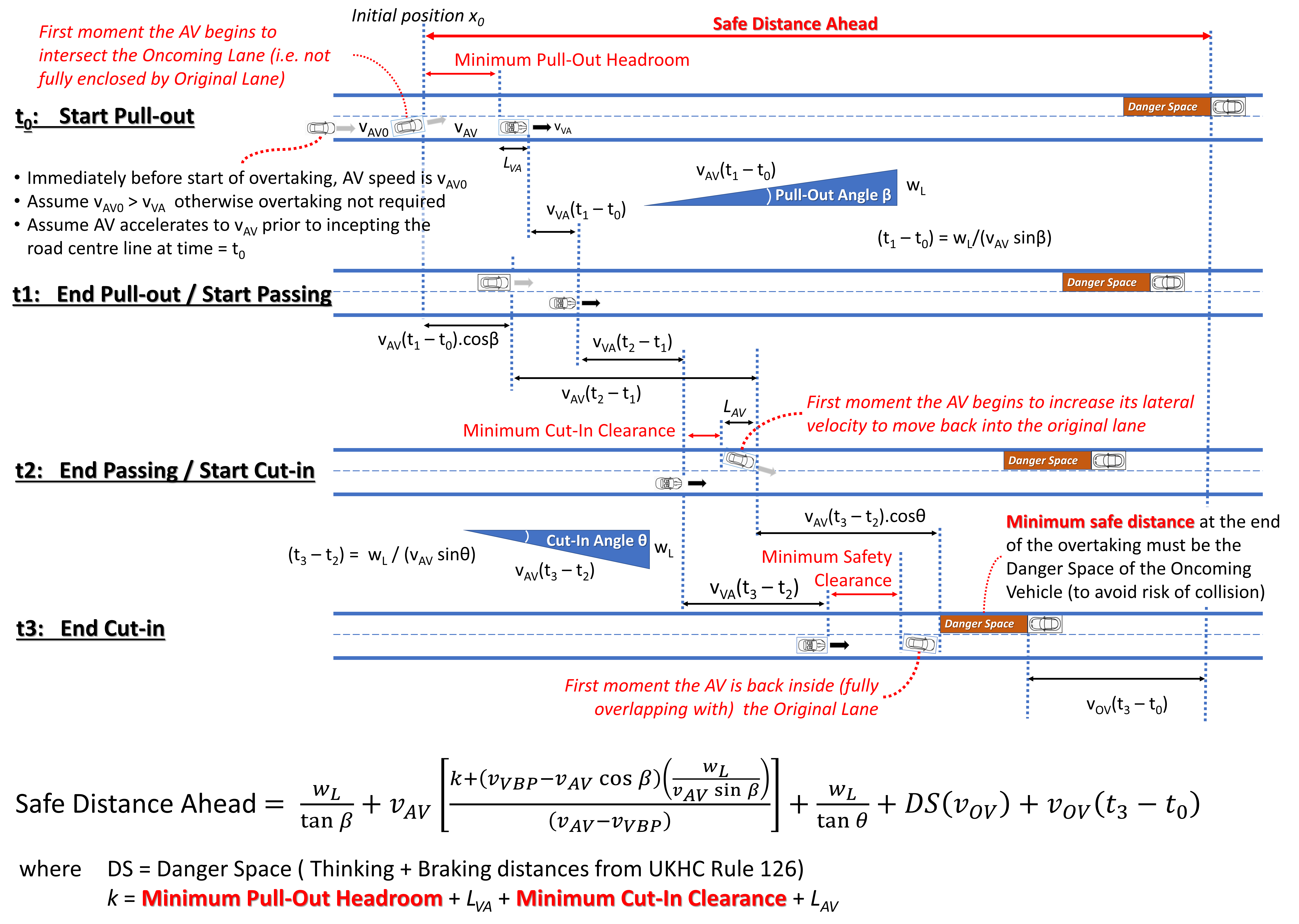}
    \caption{Schematic of Model-based Analysis of Overtaking Manoeuvre.}
    \label{fig:overtaking_analysis_diag}
\end{figure*}
\endgroup

\end{document}